# Exploring structure-property relationship on a nanoscale for tailoring films of amphiphilic polymer co-networks

Kevin Hagmann[1], Carina Schneider[1], Stephanie Ihmann[2], Frank Böhme[2], Regine von Klitzing[1]

[1]Institute for Condensed Matter Physics, Technische Universität Darmstadt, Hochschulstr. 8, D-64289 Darmstadt, Germany

[2]Leibniz-Institut für Polymerforschung Dresden e.V., Hohe Straße 6, D-01069 Dresden

**Abstract**

Amphiphilic polymer co-networks (APCNs) provide a large toolbox for tuning coatings important for applications such as bio-interfaces. Therefore, we investigate the influence of network composition and environmental conditions on the structure and mechanical and adhesive properties of thin films composed of hydrophobic tetra-PCL and hydrophilic tetra-PEG stars of varying sizes. State-of-the-art atomic force microscopy (AFM) techniques, including phase imaging, fast quantitative static indentation and dynamic indentation, provide insights into the structure-property-relationship on various length scales. PEG-rich networks exhibit amorphous morphologies with spherical nanodomains and elastic moduli of a few MPa, while PCL-rich networks form semicrystalline cylindrical arrangements with moduli up to several hundred MPa in water. Temperature-dependent measurements in water revealed a strong hysteresis of elastic moduli while shifting the melting/crystallization transitions or preventing crystallization in PEG-rich networks. All networks displayed predominantly elastic behavior. Co-networks in non-selective solvent conditions are overall softer, less adhesive and structurally more homogeneous. These results establish a predictable correlation of network composition, physical and chemical environment, structure and properties, which makes them suitable for a rational design of amphiphilic systems for various applications.

**Introduction**

Amphiphilic polymer co-networks (APCNs) combine hydrophilic and hydrophobic building blocks.[1,2] The classification as a co-network implies that it consists of at least two distinct polymer types that are linked together physically or chemically, where each polymer provides its own identity in terms of monomer composition and sequence. It distinguishes itself from copolymer networks, where instead different monomer units are built into the same polymer chain and crosslinked afterwards. Polymer co-networks, therefore, provide their own set of features that can be fine-tuned by the chosen polymer type. When choosing a hydrophilic polymer and crosslinking it with a hydrophobic polymer, the resulting 3-dimensional co-network is amphiphilic. The network then shows a very complex interaction with solvent systems of varying polarity, allowing it to swell in both aqueous and organic solvents.[3,4] Likewise, one or more polymer types within the co-network can collapse into various structural architectures if the solvent affinity is low. In this context, it is important to distinguish between solvents that are able to solely swell one polymer type in the co-network and solvents that allow all polymers to swell. If a solvent only swells one or a limited number of polymers in the co-network, while the other polymer types do not swell or even collapse further, it is classified as a selective solvent. For amphiphilic polymers, such solvents are usually found at the edges of either very high or very low polarity. In contrast to selective solvents, non-selective solvents can swell all polymer types in a network. They may also be referred to as cosolvents. The ambiguous swelling behavior of APCNs provides extensive academic and industrial opportunities. Various network properties, such as network permeability, transport, and elasticity, can be controlled in this manner and made suitable for various applications. Materials that require the simultaneous interaction of hydrophilic and hydrophobic substances serve as prime examples for the use of APCNs. In terms of industrial applications, soft contact lenses are the main beneficiary of amphiphilic networks.[5–7] Transport of ions in aqueous electrolytes (hydrophilic) as well as permeability of oxygen (hydrophobic) are crucial requirements for such materials. Other applications include membrane systems for fuel cells, antimicrobial coatings, sensors, and cell culture topics.[8–10] A special interest is also put on drug delivery systems.[11–14] In each of these systems, APCNs provide interfaces and bulk material whose compatibility and properties can be controlled and switched, depending on polymer composition and environmental conditions.

APCNs provide a great number of opportunities, but their formation is very complex. In order to design high-performance materials, a number of questions arise. What polymers should be used to form an amphiphilic network, and what monomers are used to synthesize these polymers? How should the polymer structure be designed, and what type of connectivity should be used? How will the synthesis conditions and environmental conditions after the synthesis impact the resulting network structure and its subsequent properties? What types of defects will occur? To answer these questions, it is useful to start from model systems where each step of the way can be observed and controlled; however, their synthesis can be difficult.[1,15,16]

It has been demonstrated by Sakai et al. that polymer co-networks can be designed with reduced network defects when coupling tetra-arm star polymers.[17] Model networks with hetero-complementary coupling of functionalized A4- and B4-type tetra-PEG stars were synthesized. Previous works highlight the use of PEG in hydrophilic networks as well as other star-type polymers, consisting of three-arm polymer stars, which is the minimum number of arms required for a 3-dimensional network.[18] The advantage of star-shaped polymers lies in the fact that network mesh sizes are somewhat predetermined by precursor size and shape. Furthermore, network connectivities can be navigated by initial synthesis conditions. In order to create homogeneous networks while maintaining uniform amphiphilic behavior, the hydrophilic and hydrophobic polymer building blocks need to have different reactive groups. When designing each polymer star type of varying hydrophobicity with different terminal reactive groups and connecting them with a bifunctional coupling agent, homogenous co-networks can be formed. This leads to a 3-dimensional network whose hydrophilic and hydrophobic polymer stars are always alternating. By choosing the appropriate bifunctional linker and different terminal groups for the polymer stars, it is impossible to create inhomogeneities through connections of the same star type, avoiding some of the most common defects in polymer networks.[16,19] Network defects are then limited to unreacted groups and multiple-arm connections. Through this approach, the network architecture is somewhat predictable and avoids common defects. Scattering experiments,[20–22] low-field nuclear magnetic resonance (NMR) measurements[23,24], and computer simulations[19] have demonstrated that network inhomogeneities are rare. However, there is some degree of connectivity error, which occurs mainly in double-arm connections. Nonetheless, the overall network structures are reasonably predictable, allowing the modelling of properties, such as mechanics, where each chain contributes to the overall picture. Persisting defects include finite loops and dangling chains that still influence the mechanical properties.[25]

Recent works by Bunk et al.[26] built upon the synthesis strategy of Sakai et al. by expanding the co-networks' variability through the introduction of a second star type with opposing hydrophobicity. Amine-terminated tetra-arm poly(ethylene glycol) (tetra-PEG-NH$_2$) serves as the hydrophilic building block, whereas 2-(4-nitrophenyl)-benzoxazinone-terminated tetra-arm poly(ε-caprolactone) (tetra-PCL-Ox) acts as the hydrophobic building block. The hydrophobic star already contains the bifunctional coupling agent. Tetra-PCL-OH is therefore reacted with benzoxazinone in a first step. The other reaction site of the coupling agent is then available only for the amine-terminated tetra-PEG-NH$_2$. In this manner, only alternating hydrophobic/hydrophilic star connections are possible. The reaction scheme is displayed in Figure 1. The coupling plays a crucial role in creating homogeneous networks. Not only does it guarantee an interchangeably connected network through chemical selectivity, but it also allows for control of the reaction kinetics.

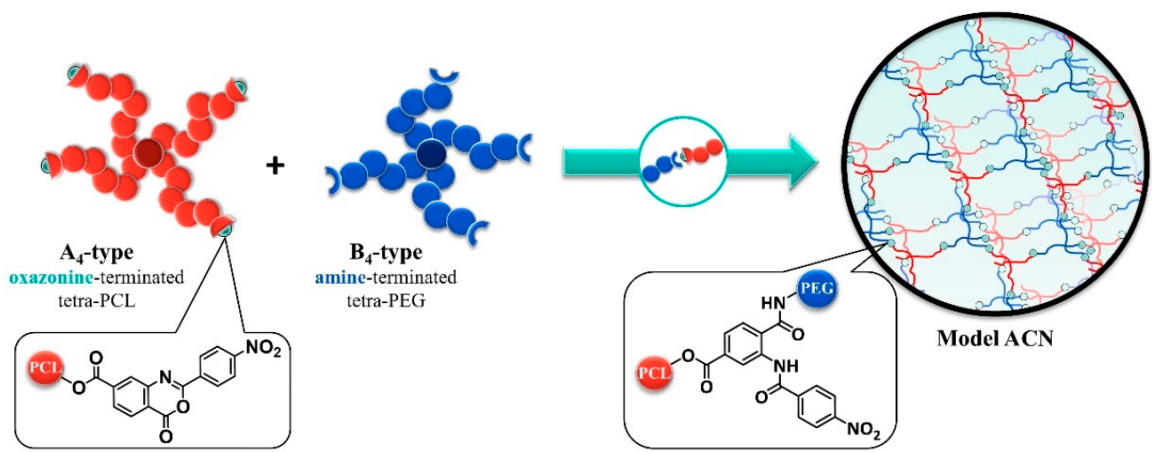

**Figure 1:** Formation of APCN co-network from tetra-PCL and tetra-PEG through hetero-complementary coupling. Adapted with permission.[26,27]

The reaction is performed in a non-selective solvent to allow for sufficient mixing, resulting in the formation of a polymer gel. Different preparation techniques enable the formation of bulk gels of mm- to cm size, or thin films in the nm-μm range. The preparation of the latter lends itself to not only studying the bulk properties of the material but also focusing on the interface/surface of the gel material.

*Surface analysis with atomic force microscopy (AFM)*

For applications in which interface properties play a role, such as targeted adhesion or coatings, both bulk properties and interface properties must be analysed. For the analysis of structures and properties at the surface of gel films, atomic force microscopy is particularly suited and is used as the main tool in the presented studies. Depending on the research question, atomic force microscopy (AFM) can provide valuable information in addition to common surface-sensitive tools like X-ray scattering or neutron scattering, which have been used to study thin films of polymers effectively.[28–30] Most notably, AFM provides real-space images and doesn't require complex models to obtain information on structural features.[31] Other powerful techniques like electron microscopy often require more complex sample preparation and extreme measuring conditions (e.g. vacuum). AFM shines with its ease of setup, use and compatibility with various environmental conditions. Another great advantage of AFM techniques is the determination of not only structural information but also information on surface properties, such as elastic modulus and adhesion. Recent advancements in AFM technology allow for much faster image acquisition, while simultaneously monitoring multiple channels representing surface properties. Since different indenter types and geometries can be used with this technique, AFM can provide answers to questions on various length scales and orders of magnitude in terms of physical properties. In the following studies, AFM is used as the central tool for gathering information on structure-property relationships. The most important imaging techniques are highlighted here:

In AFM tapping mode, cantilever oscillations near its resonance frequency are induced by a piezoelectric crystal while the cantilever is moved across the sample surface. It intermittently touches the sample surface, and due to tip-sample interactions, the amplitude of the cantilever oscillation is damped while simultaneously shifting the phase of the induced oscillation. Through the damping of the amplitude, tip-sample interactions can be monitored and minimized to prevent tip and sample damage. Information on sample height can be obtained via height topography images. The phase change of the oscillation is a result of the physical and chemical interaction that can be attractive or repulsive, and therefore provides a second channel of information that can be displayed as a phase image. It does not directly provide quantitative information on sample properties, but qualitatively provides superior contrast to other imaging techniques.

Advanced imaging techniques are available to extract sample properties simultaneously with structural information. This is done through AFM indentation experiments (Figure 2A). A cantilever approaches the sample surface and reaches the contact point. It then indents the sample up to a given target force, after which it retracts fully and returns to its starting height. Modern AFM techniques were used in this work, e.g. QI-Advanced (QI: Quantitative Imaging), which have optimized approach and retraction cycles, with cantilever speeds ranging between 50-100 µm/s, allowing for hundreds of indentation cycles per second. This results in structure-property acquisition with high resolution on time scales in the same order of magnitude as classical tapping modes. Approach curves can be used to generate height images by identifying the contact point. In the following studies, approach curves were also used to determine elastic moduli after fitting data with the Hertz model.[32] Sample stiffness was calculated using tangential fits of the approach curve at a given setforce, denoted by $F_S$. The retraction curve was used to identify the maximum adhesion force ($F_{adh}$) that the cantilever receives from a sample before it snaps back out of contact.

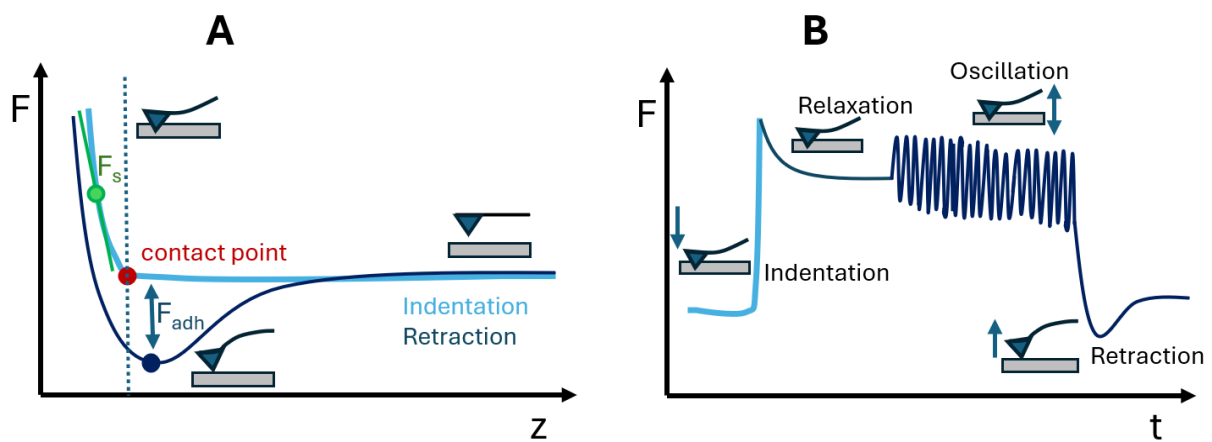

**Figure 2:** AFM - Scheme of Static (A) and dynamic (B) indentation measurements. (A) Force vs distance curve (F-z). Cantilever approaches the sample until the contact point, followed by indentation. Retraction of the cantilever to the original position. $F_{adh}$ = Maximum adhesion force. $F_S$ = Setforce for tangential slope fit. (B) Force vs time curve (F-t). Cantilever approach and indentation followed by relaxation. Oscillation with set frequency, amplitude and duration, followed by cantilever retraction to the original position.

Dynamic indentation modes are utilized to acquire viscoelastic information at the surface of a sample (Figure 2B). Approach procedures are similar to static indentation measurements, but after a cantilever indents a sample to a given target force, it is left in place to allow for sample equilibration, after which an oscillation in the range of 1 – 500 Hz can be applied for a few seconds with the cantilever tip residing in the sample. The resulting damping of the oscillation amplitude, as well as the phase shift of the oscillation, is utilized to derive storage and loss moduli. Afterwards, the cantilever retracts to its original position before repeating the cycle. This measuring technique is an extension of the static indentation experiments described above. It provides additional sample property information such as viscoelasticity, but due to the additional time requirements for equilibration and oscillation, its drawbacks include either longer measuring times or lower resolution.

The studies listed below demonstrate the application of classical scanning and indentation experiments, while newly reported data in the results and discussion sections encompass recently developed, fast quantitative imaging techniques as well as dynamic modes.

*Interfacial properties of APCN-gel films based on tetra-PEG and tetra-PCL[27]*

The studies presented here focused on the synthesis of polymer gel films, both in bulk and thin films, from tetra-PEG-$NH_2$ and tetra-PCL-Ox precursors onto a suitable substrate. Reaction conditions and sample treatment were optimized to gain access to surface analytical tools. AFM was used to investigate the structure and mechanical properties of gel films in various environmental conditions.

In a first step, polymer gel films with planar surfaces were prepared to study the structure and properties at the surface. The reaction is analogous to the procedures reported by Bunk et al., for the synthesis of bulk material[26] but tweaked towards special requirements for thin films.[27] The initial step is a bulk reaction of the precursors tetra-PEG-$NH_2$ ($M_w$ = 10kDa) with tetra-PCL-Ox ($M_w$ = 10kDa) in the non-selective solvent toluene at 60°C. Under these conditions in the bulk, the reaction duration is approximately one hour, as confirmed by NMR studies.[26] The resulting gels are, however, difficult to handle due to drying and reswelling, resulting in macroscopic deformation of the networks, which causes the gel to delaminate from its underlying substrate.

This delamination makes subsequent sample preparation for surface analysis with atomic force microscopy very challenging. To overcome this problem, thin gel films were prepared with an adapted procedure: The reaction in bulk is allowed to proceed close to its halfway point, at which the network is partially formed, but has not reached the gel point[26,33], where it would be difficult to process further. By spin casting[34,35] the viscous but still processable reaction mixture onto silicon substrates, thin films with a thickness of 300 – 1000 nm were obtained. The reaction is then finished in the melted state at a temperature of 60°C. The resulting thin films have smooth surface structures with a surface roughness of 2 – 8 nm in ambient conditions. Sample deformation and delamination are prevented by covalently linking the polymer network onto the substrates, using amino-silane chemistry. Before spin casting, cleaned and activated silicon substrates are treated with 3-aminopropyl-triethoxysilane (APTES) to cover the sample surface with amino-groups.[27,36,37] The benzoxazinone coupling agent in the network can then covalently link to these amino groups in the same manner as it would find a tetra-PEG-$NH_2$ partner. The resulting gel films were stable and did not delaminate even under multiple solvent exchanges. Thick "bulk" films were prepared by pouring the pre-reacted solution on APTES-coated substrates and served as bulk reference samples.

Subsequent AFM measurements in tapping mode enable the structural investigation of film surfaces in ambient, aqueous and organic solvent conditions (Figure 3). The surface topology of PCL_PEG films appeared mostly structureless and amorphous on a microscopic level, exhibiting surface roughness of 7 – 35 nm and 17 – 66 nm for thin films and bulk films. Ambient conditions displayed the lowest surface roughness, whereas swelling of samples in toluene or water led to similarly increased surface roughness. Similarities in the observed surface structures between thin films and bulk films for all environmental conditions validated the previously mentioned synthesis protocols. While the resulting structures of both sample types were similar, the mechanical properties appeared vastly different.

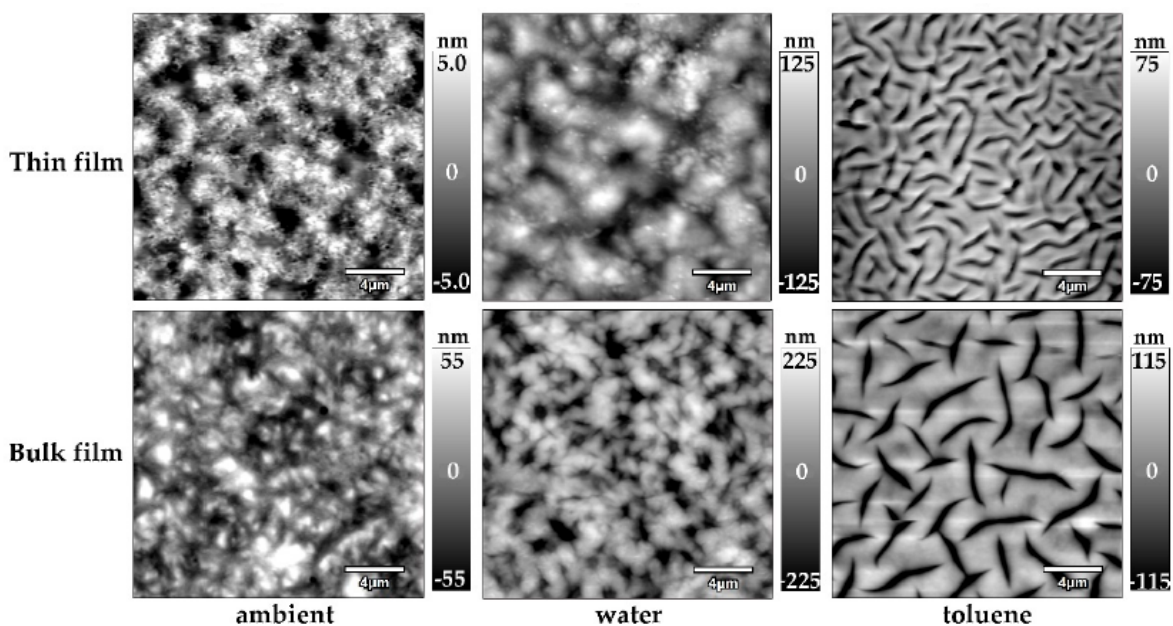

**Figure 3:** AFM surface topographies of gel films made from *tetra-PEG-NH$_2$ (M$_w$ = 10kDa) and tetra-PCL-Ox (M$_w$ = 10kDa)* in various environmental conditions. Ambient (left), swollen in water (middle), swollen in toluene (right). The top row shows thin films. The bottom row shows bulk film samples. Scale bar = 4 μm. Adapted with permission.[27]

This was confirmed by static AFM indentation experiments in selective and non-selective solvent conditions (Figure 4). Thin films swollen in water exhibited ten times higher elastic moduli than corresponding bulk samples. This is attributed to the synthesis procedure where spin-casting of an unfinished network leads to different levels of the network interconnecting and increasing tightness, which in turn decreases swelling ability. A similar trend is observed for samples swollen in the non-selective solvent toluene. Thin films are significantly harder than bulk samples. Overall, samples swollen in toluene exhibit elastic moduli lower than those in water. In a non-selective solvent, both network components can swell, creating an overall soft network, whereas the selective solvent water only partially swells the tetra-PEG stars, significantly increasing network stiffness.

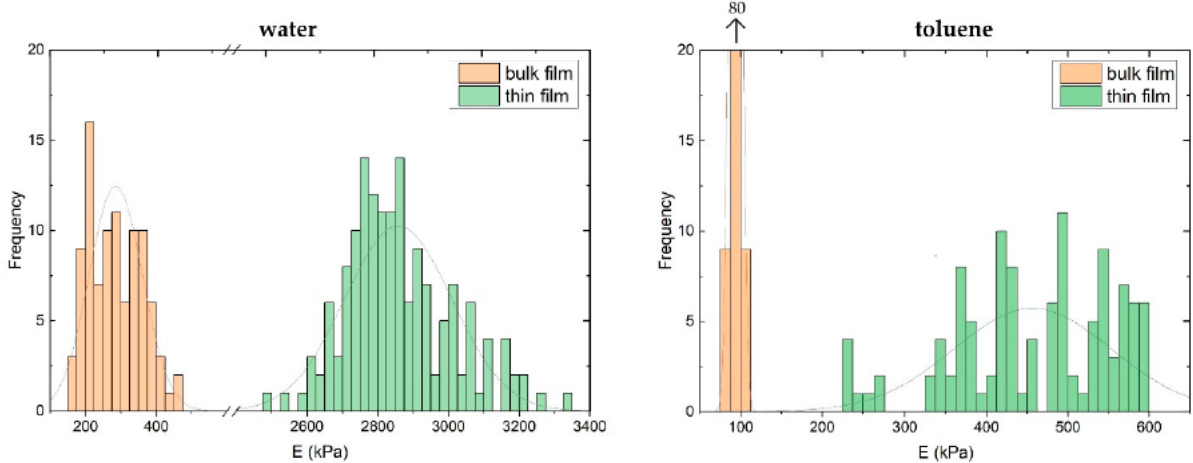

**Figure 4:** AFM-indentation experiments with tip (tip radius < 30 nm). Elastic moduli distribution of thin gel films (green) and bulk films (orange), swollen in either the selective solvent water (left) or non-selective solvent toluene (right).[27] The films were made from tetra-PEG-NH$_2$ (M$_w$ = 10kDa) and tetra-PCL-Ox (M$_w$ = 10kDa).

Structural investigations with advanced AFM techniques, such as phase imaging, revealed the presence of nanoscopic spherical/elongated structures, which indicate phase separation. Phase separation is expected in a system that contains both hydrophilic and hydrophobic polymers. While non-linked polymers will separate macroscopically into different phases when subjected to a selective solvent, the alternating star-architecture in these APCNs prevents this mode of action. Instead, the system undergoes phase separation on a nanoscopic scale.

*Impact of swelling and structure on macroscopic and nanoscopic mechanical properties of APCNs in selective and non-selective solvents*[38]

To further study the structural phase separation on a nanoscopic scale in different chemical environments and potentially correlate the structure with its mechanical properties, bulk films were further investigated on different length scales using atomic force microscopy with different indenter geometries and compared to macroscopic rheometry measurements (Figure 5).

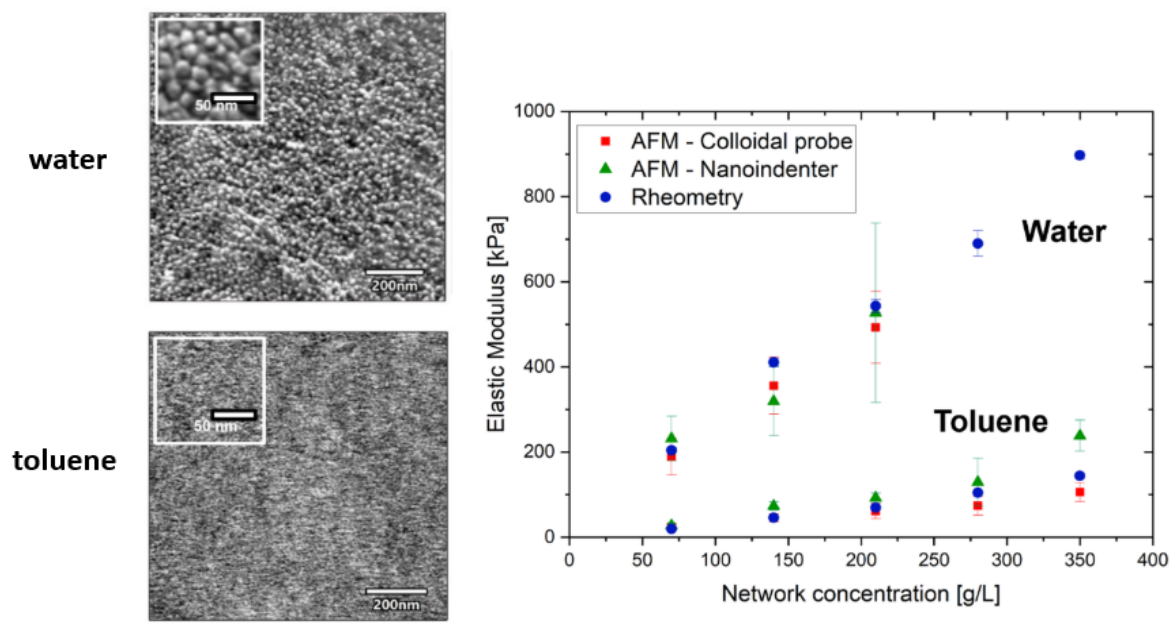

**Figure 5:** Elastic moduli of thick "bulk" films in dependence of precursor concentration during network formation in the selective solvent water and non-selective solvent toluene obtained from different measuring techniques (AFM, rheometry) on different length scales (right). Correlation with spherical nanophase separation in water (top-left) and lack of structures in toluene (bottom-left). Data taken from[38]. The bulk films were made from tetra-PEG-NH$_2$ (M$_w$ = 10kDa) and tetra-PCL-Ox (M$_w$ = 10kDa).

In the selective solvent water, spherical domains with a size of approximately 20 nm appear at the surface. Similar domain sizes have been found via small-angle X-ray scattering (SAXS) measurements with domain distances of 19 nm and radii of 6 nm, whereas these domains were also found in Monte-Carlo-Simulations and attributed to the aggregation of 17-25 PCL stars.[39] This aggregation and exclusion of water from the inside of PCL clusters noticeably reduces the degree of swelling of the co-network and should have a profound impact on the mechanical properties. This was confirmed by the elastic moduli of the bulk films being in the range of a few hundred kPa that scale linearly with concentration. In contrast, samples in toluene had elastic moduli in the range of tens of kPa, or an order of magnitude lower. In the case of toluene, this suggests a fully swollen network, which was further evidenced by a higher degree of swelling and a completely homogenous surface with no visible structures. Overall, results from the investigation on various length scales suggest that structure and mechanical properties are similar at the bulk and surface of APCNs in this study,[38] which has proven to be important for amphiphilic systems in previous studies.[40,41]

*Thin film APCN coatings with enhanced cell adhesion*[42]

The adaptability of these amphiphilic co-networks with controlled structural elements and mechanical properties makes them potentially suitable for bio-based drug carrier and release systems. Thin PEG-PCL films were therefore used in cell adhesion studies.[42] The parameter space was broadened by including co-networks that are formed from tetra-PCL and tetra-PEG building blocks with varying star size and varying concentrations during network formation. Standard networks with 47 wt-% PEG have not demonstrated cell viability by themselves. However, this could be improved when either the PCL content of the network is increased to at least 85%, or the networks with higher PEG content are surface modified with extracellular matrix. The variation of initial reaction concentrations and star size had a profound impact on the networks' swellability in selective and non-selective solvents as well as the mechanical properties in cell medium, which impacted cell viability. Surface structures in aqueous conditions displayed vastly different microstructures. Most notably, the presence of spherulites was observed in samples with higher PCL content, which influenced the networks' ability to transport hydrophobic or hydrophilic dyes that were used to simulate drug uptake and release.

Key results from studies in this project include the successful immobilization and synthesis of thin, smooth APCN gel films while maintaining structural similarity to bulk samples. Investigations of nanomechanical and micromechanical properties have demonstrated that APCN films can be designed with a wide range of stiffnesses. The relationship between the degree of swelling in selective and non-selective solvent environments and the mechanical properties of bulk APCN gels has also been explored. The implications for biological applications have been demonstrated by tailoring network composition and surface modifications to various outcomes on structure and properties, and subsequently cell viability, drug release and uptake.[27,38,42]

While the aforementioned studies describe a pathway towards designing materials that have controlled structures and properties with potentially predictable use cases for bio-applications, they still miss key features. Although progress was made to correlate surface structures to their properties, studies still do not fully allow for a direct structure-property-evaluation since structural information was obtained separately from its property information. Mechanics have been probed, but they present sample averages and cannot be directly tied to a local environment. This, however, is critical since the presence of phase-separated domains in the nano-regime demands a more precise correlation with their local mechanical properties. Bio-application tests suggest that cell viability and drug uptake and release are strongly tied to the nanostructure and nanoproperties, but this was mostly demonstrated for APCNs with high PCL content, whose nanostructure or type of phase separation is still unknown.[42] The presence of semi-crystallinity in bio-conditions (aqueous, 37°C) on a microscopic scale also raises questions on the precise nanostructure and properties over a wider range of temperatures, while also considering crystallization dynamics. This would highlight potential differences during the production of the APCNs, their storage and processing history. Furthermore, cell viability studies also highlighted that cell adhesion or surface modification critically depend on mechanical properties, chemical environment and hydrophobicity.[42] Adhesive properties of APCN have not been considered, but may prove critical. Viscoelastic properties have been neglected as well, although they may deliver some key insights into differences in cell activity after proliferation.

The following study attempts to directly correlate structural and property information down to the nanoscale using state-of-the-art atomic force microscopy techniques. Elastic Moduli, network stiffness and adhesion of co-networks with varying star size shall be analyzed nanoscopically in direct comparison to their corresponding phase separation in selective and non-selective solvents. Deformation-rate-dependent viscoelelastic properties of gel films shall be considered to determine a key parameter for cell activity. Environmental conditions are expanded by studying structural and property relations for APCNs in selective and non-selective solvents as well as temperature cycles.

The goal of this study is to highlight a direct pathway and control to design an amphiphilic material that has a predictable structure-property-relationship under various environmental conditions and develop a strategy that can be immediately transferred to produce gel films for applications such as drug uptake and delivery.

**Experimental section**

*Materials and preparation:*

*Polymer synthesis:* All reagents and solvents were of analytical grade and obtained from Sigma-Aldrich unless otherwise specified. Hydroxyl-functionalized tetra-arm PEG was purchased from JenKem Technology (Plano, USA) . ε-Caprolactone (ε-CL) was dried over $CaH_2$ under reduced pressure for at least 24 h, distilled under vacuum, and stored under nitrogen. Tin(II)-2-ethylhexanoate [$Sn(Oct)_2$] was similarly purified by vacuum distillation and stored under inert atmosphere.

Tetra-arm PEG-$NH_2$ and tetra-arm PCL-Ox were synthesized following established protocols by Bunk et al.[26]. Briefly, tetra-arm PEG-$NH_2$ was obtained via reductive amination of hydroxyl-functionalized tetra-arm PEG, and tetra-arm PCL-Ox was synthesized using oxazinone coupling chemistry with hydroxyl-terminated tetra-arm PCL. All polymers were characterized by $^1$H-NMR spectroscopy to confirm molecular weight and functionalization efficiency. The molecular weight $M_w$ of the PEG stars was 2, 5 or 10 kDa, called PEG2, PEG5, PEG10 in the following. PCL stars had a $M_w$ of 10 or 20 kDa, called PCL10 and PCL20, respectively, in the following.

*Wafers*: Silicon wafers were purchased from Silchem GmbH (Freiburg, Germany) and cut into 10 x 10 mm squares using a glass cutter. (3-Aminopropyl)triethoxysilane (99%, APTES). Hydrogen peroxide (30%) was purchased from VWR BDH chemicals (Pennsylvania, United States). Sulfuric acid ($\geqslant$ 95%) and toluene ($\geqslant$ 99.8%) were purchased from Fisher scientific UK (Loughborough, United Kingdom). Extra-dry toluene (99.85%) was acquired from Thermo Fisher Scientific (Waltham, MA, United States). Ultrapure water (resistivity $\geqslant$ 18.2 MΩ·cm, 25°C) was prepared using a Merck Millipore™ Milli-Q system.

The Silicon wafers (10 x 10 mm) were cleaned sequentially in acetone, ethanol, and Milli-Q water. Clean substrates were etched in piranha solution (3:1 $H_2SO_4$:$H_2O_2$) for 20 minutes, rinsed with Milli-Q water, and dried under a stream of nitrogen. Afterwards, the wafers were coated with APTES.

*Thin Film preparation*

Thin films of tetra-PCL_tetra-PEG co-networks were synthesized and immobilized on silicon substrates based on established procedures.[26,27] as described above in the introduction. The present study focuses on combinations of different star precursor sizes: PCL10_PEG2, PCL20_PEG5, PCL20_PEG10, and PCL10_PEG10. One size of PEG stars was mixed with one size of PCL stars at an equimolecular ratio. This procedure enables the variation of the PEG/PCL ratio at an equimolar number of connection points between the different stars. The total concentration of stars was 70 g/l in toluene (non-selective good solvent). After a reaction time of 20 min at 60°C, the solution was spin-coated at 5000 rpm for 1 minute on a Silicon substrate pre-coated with APTES. All samples were stored in ambient conditions before measurements with atomic force microscopy. The combination of different sizes of PEG and PCL stars leads to a variation of the PEG/PCL ratio: PCL10_PEG2 (9 wt-% PEG), PCL20_PEG5 (15 wt-% PEG), PCL20_PEG10 (38 wt-% PEG) and PCL10_PEG10 (47 wt-% PEG).

*Atomic Force Microscopy (AFM): topography and mechanical/rheological properties*

AFM measurements were performed on thin films swollen in either water or toluene using a variety of AFM modes:

The surface topography in water was recorded in tapping mode at 25 °C while monitoring height and phase channels. The measurements were performed with a Cypher AFM (Oxford Instruments, Wiesbaden, Germany) utilizing the cantilevers AC240TSA (70 kHz, 2 N/m, tip radius 7 nm) and HiRes-C14/Cr-Au (160 kHz, 5 N/m, tip radius < 1 nm) from MikroMasch®.

Indentation-based experiments for structure, elastic modulus, stiffness and adhesion were performed on a Nanowizard IV AFM (Bruker, Berlin, Germany) in QI-advanced mode with tens of thousands of indentations per force map, while applying a force of 10 nN for indentation. NSC18 Cr-AU/BS cantilever (65 kHz, 2.8 N/m, tip radius < 8 nm) from MikroMasch® was used for all indentation-based measurements. The resulting force-distance curves were batch-processed using built-in software features of Digital Processing 8.0 (Bruker, Germany). After calibration and background subtraction/baseline corrections, approach curves were fitted with the Hertz-Model up to an indentation force of 10 nN, considering indenter geometry and a Poisson-ratio of 0.5 to obtain elastic moduli images and distributions.[21,43] Approach curves were additionally fitted tangentially at the maximum setforce of 10 nN to obtain stiffness values. Retraction curves were used to extract the maximum adhesion force. The total number of force curves analyzed for each sample type and sample property exceeds 100,000.

Viscoelastic properties were similarly obtained using self-programmed measurement profiles. Oscillations ranged from 10 – 250 Hz in logarithmic steps. The applied setforces were 5, 10, and 20 nN for water measurements, and 1, 2, and 5 nN for toluene measurements. After each indentation and measurement, samples were allowed to equilibrate for at least 0.5 s, which was sufficient to create stable baselines during oscillation. Results from viscoelastic measurements were batch-processed using built-in software features of Digital Processing 8.0 to obtain storage moduli, loss moduli and loss tangents.

**Results**

The surface of PCL10_PEG2-, PCL20_PEG5-, PCL20_PEG10- and PCL10_PEG10 films was analyzed by AFM with tapping mode in the selective solvent water on small length scales (1 x 1 µm$^2$). Height and phase channels are displayed in Figure 6. Height imaging reveals the surface topography of a sample. However, the roughness of the sample can prevent the visualization of potential underlying nanoscopic structures, which is further complicated by a tip radius of curvature in the same order of magnitude as the expected structural features (tip radius of curvature = 7 nm). Samples with high PCL amount show some areas with preferred orientation, but further identification remains difficult due to surface roughness. Phase imaging (Figure 6, bottom row), therefore, provides a better opportunity to visualize structural features. Surface roughness is eliminated, and different physical and chemical interactions of the cantilever and sample are seemingly differentiated on a length scale smaller than the radius of curvature of the tip. Samples with higher amounts of tetra-PCL (PCL10_PEG 2, PCL20_PEG5 and PCL20_PEG10) show spherical and/or cylindrical nanoscopic structures with a preferred orientation. This behavior has been suspected previously when looking at much larger length scales, where these types of networks showed µm-sized semi-crystalline structures in the form of spherulites.[42] Only the PCL10_PEG10 film does not show any preferential orientation (Figure 6, micrographs at the outer right), which has been recognized before (see Figure 3, top middle, Figure 5, top left). Beyond some undefined structuring, only spherical phase separation can be observed, which was previously highlighted for bulk samples.[38] This is a further proof that the suggested synthesis protocol for thin films results in structural features similar to those that would be present in a bulk sample.

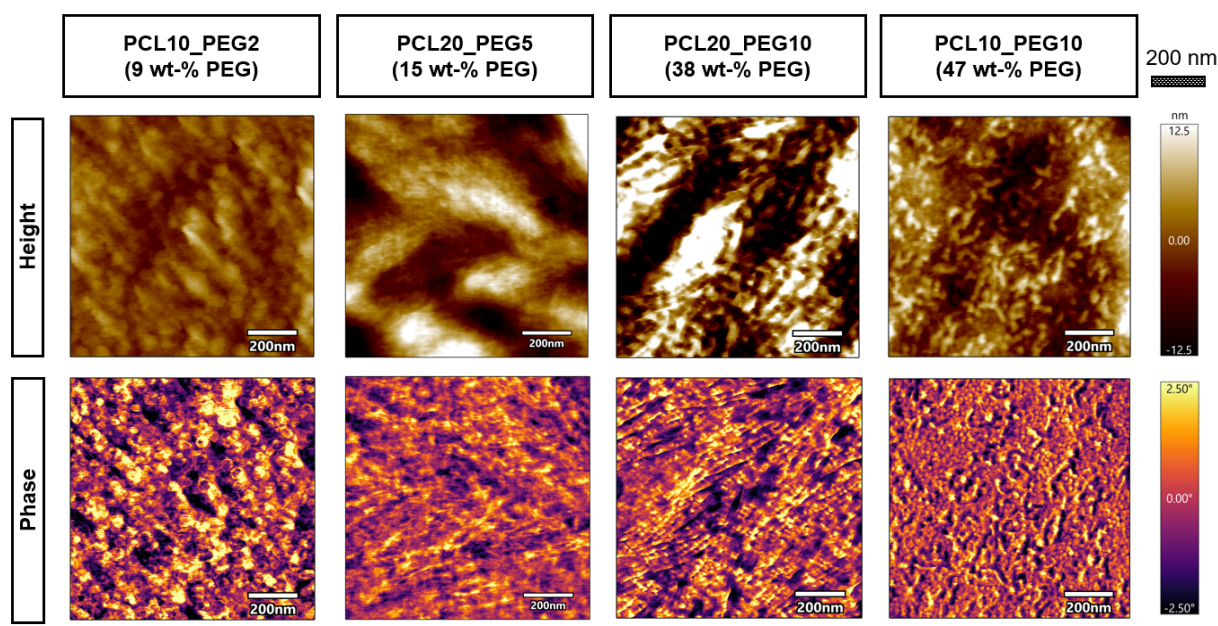

**Figure 5:** Surface topographies of PCL_PEG films in water measured by AFM. Height (top row) and phase (bottom row) channels from tapping mode. Image size = 1 x 1 μm$^2$. Scale bar = 200 nm.

The resolution of height imaging in water was further improved in AFM tapping mode through the use of cantilevers with < 1 nm tip size (Figure 7). Representative samples of PCL10_PEG2 and PCL20_PEG10 were used to highlight the significant improvement of height images when the cantilever size is significantly smaller than the investigated structures. Figure 7 displays a comparison between experiments with different cantilever sizes. Height images now serve as direct proof for the presence of cylindrical/spherical structures in PCL10_PEG2, as well as strongly oriented cylindrical structures across the entire image for PCL20_PEG10. Sharp tips allow for a more accurate quantification of surface structures. Samples displayed in Figure 7, PCL10_PEG2 and PCL20_PEG10, have a surface roughness of 5 ± 1 nm and 11 ± 2 nm, respectively. The average diameter of cylindrical fibers in PCL10_PEG2 is 22 ± 2 nm and in PCL20_PEG10 30 ± 3 nm, demonstrating that larger PCL aggregates can form with increasing tetra-PCL precursor size in the co-network.

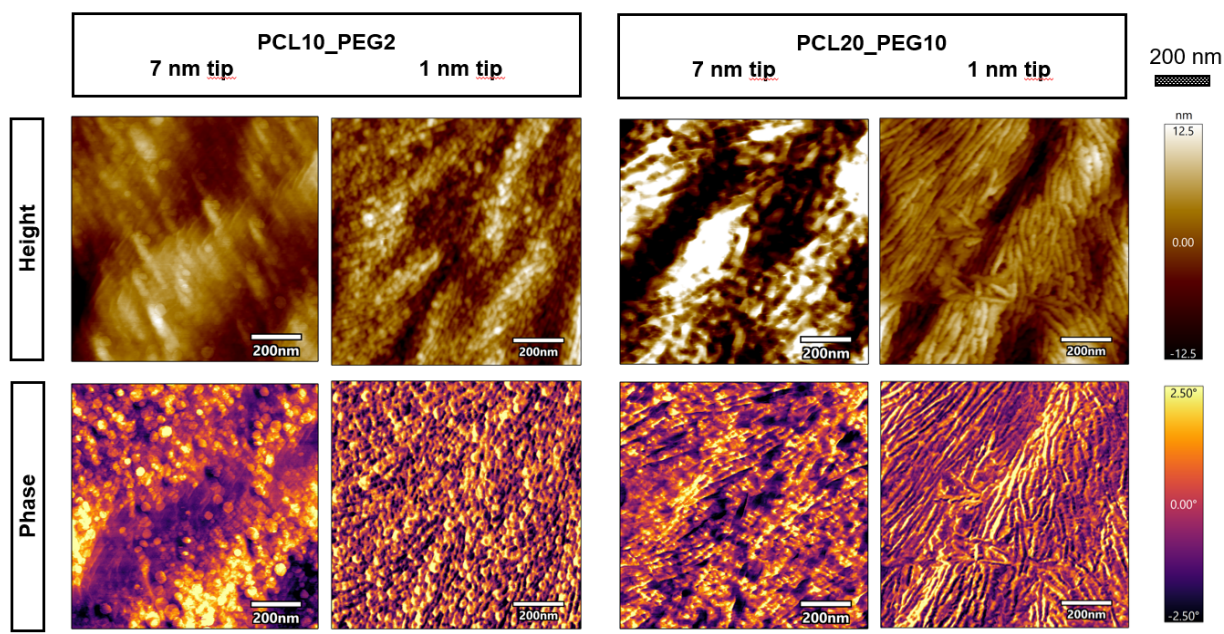

**Figure 6:** Surface topographies of selected PCL_PEG films in water measured by AFM. Height (top row) and phase (bottom row) channels from tapping mode with different cantilever sizes. Image size = 1 x 1 µm$^2$. Scale bar = 200 nm.

*Quantitative AFM-imaging of APCNs in water and toluene*

Thin films of PCL10_PEG2, PCL20_PEG5, PCL20_PEG10 and PCL10_PEG10 were further investigated by advanced AFM imaging techniques that not only consider structural parameters but also monitor properties as well. The horizon of chemical environments was expanded by analyzing samples both in the selective solvent water and in the non-selective solvent toluene. QI-indentation-based measurements simultaneously yielded information of the surface structure (height), elastic modulus, stiffness and adhesion. Figure 8 shows image data derived from these measurement modes in aqueous conditions. The image size in all cases is 1 x 1 µm$^2$ with a 512 x 512 pixel grid and the same number of individual indentations (n = 262,144). With the cantilever tip size of < 8 nm, this imaging technique allows for the study of nanoscopic structures and nanoscopic properties. Figure 8, top row displays the height topography based on the first contact point of the tip with the sample surface. The surface roughness is similar for all sample types, ranging from 5 – 15 nm. This is consistent with previous findings for water-swollen systems using AFM tapping techniques.[27,42] PCL10_PEG2 shows spherical nanostructures and an underlying µm-sized substructure with preferential orientation. On top, there are larger objects with a size of a few 10 nm. The PCL20_PEG5 film shows also an underlying µm-sized substructure with preferential orientation but without spherical nanostructures. Less objects with a size of 10 nm are detected at the surface. PCL20_PEG10 shows the presence of some circular nanostructures in an otherwise undefined microstructure, with a slight preferential orientation. PCL10_PEG10 is the only sample with no preferential orientation, but rather appears microscopically amorphous with somewhat spherical nanophase separation, as mentioned before. The second and third row in Figure 8 display the elastic modulus and stiffness, respectively. Elastic moduli images are a result of fitting each individual force curve with the Hertz model up to an indentation force of 10 nN, while stiffness images are a result of the tangential fit of each individual force curve at the indentation force of 10 nN. Both measuring techniques provide similar images, although stiffness images seem to provide higher contrast. PCL10_PEG2 now displays clearer nano- and microstructures. They appear as long (range of $\mu$m), continuous nanocylinders with a diameter of 21 ± 4 nm that align reasonably well and have a clearly identifiable preferential orientation. PCL20_PEG5 displays a very similar cylindrical nanostructure with a diameter of 19 ± 4 nm, with clear alignment and preferential orientation. PCL20_PEG10 shows shorter cylindrical nanostructures with a diameter of 31 ± 5 nm, although their individual distinction is more difficult to discern. PCL10_PEG10 still appears microscopically amorphous. The larger spherical objects, observed in the height image for PCL10_PEG2 and PCL20_PEG5 are not detected either in the elastic map or the stiffness map.

The fourth row in Figure 8 displays adhesion images. They result from plotting the maximum negative force the cantilever receives during retraction from the sample. PCL10_PEG2, PCL20_PEG5 and PCL20_PEG10 lose their sharp structural distinction, and the preferred orientation is only slightly visible. The larger spherical objects, observed in the height image for PCL10_PEG2 and PCL20_PEG5 are also detectable in the adhesion map. PCL10_PEG10 displays randomly interconnected network nanostructures.

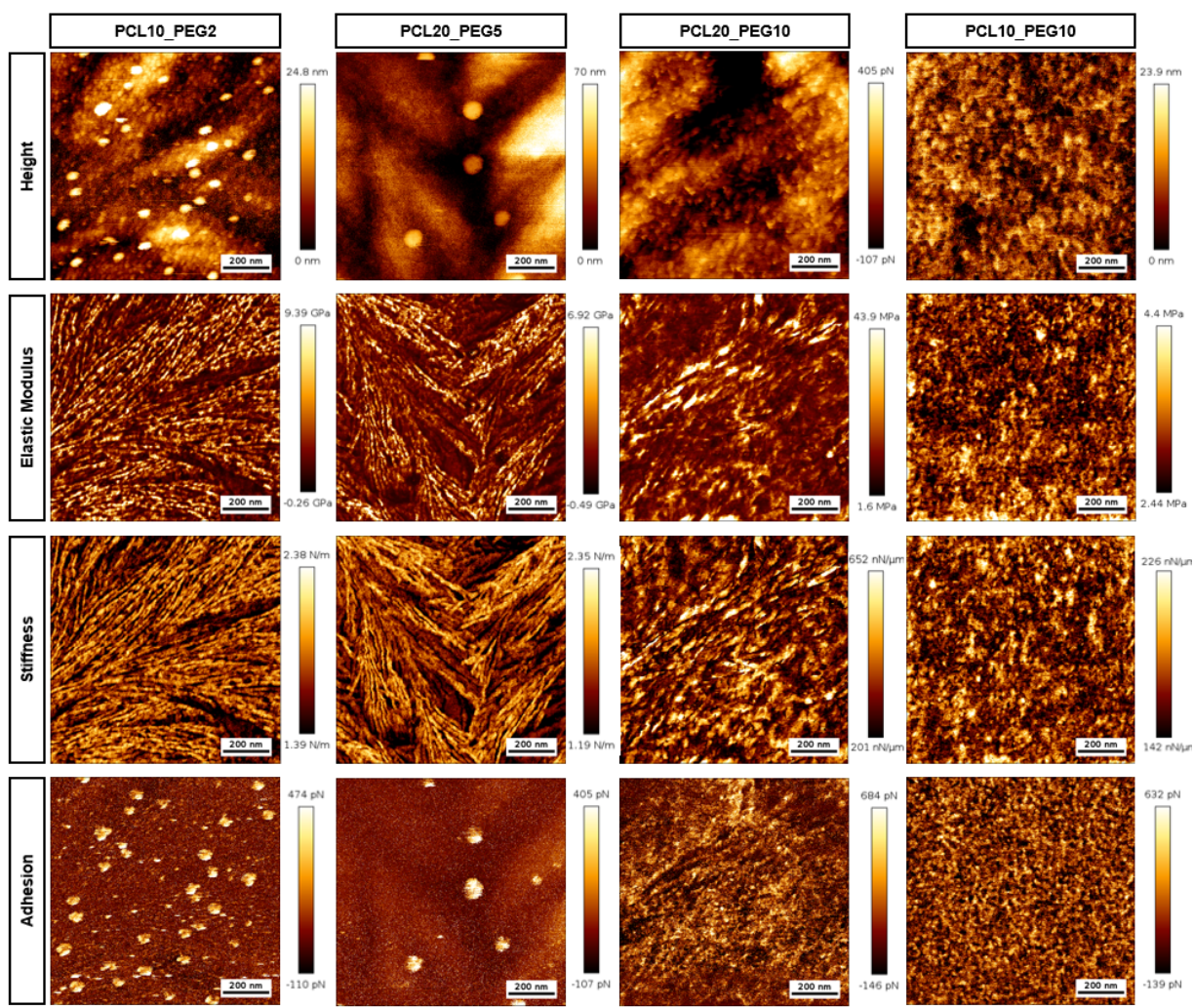

**Figure 7:** Atomic force microscopy - quantitative imaging of PCL-PEG thin films with varying star size in water. First row: Height image normalized by the contact point. Second row: Elastic moduli based on Hertz fit. Third row: Sample stiffness based on tangential fit. Fourth row: Adhesion force equal to the minimum of the force curve during cantilever retraction. Image size = 1 x 1 μm$^2$. Scale bar = 200 nm.

The imaging techniques for elastic modulus, stiffness and adhesion provide quantitative information on a nanoscale for each sample type as shown in Figure 9. Since the measurements are carried out on the nanoscale, they exhibit significant variation depending on the precise sample location. Therefore, results for all properties are plotted as a distribution. Numeric results for mean values and standard deviation are included in the supporting information Table S1. The PCL-rich films, PCL10_PEG2 and PCL20_PEG5 films, show high elastic moduli of 2.5 GPa and 1.6 GPa (likely exceeding the range of the measuring capabilities of the cantilever). In contrast, the films with a lower PCL content, PCL20_PEG10 and PCL10_PEG10 films, have significantly lower elastic moduli of 13 MPa and 3 MPa, respectively. The latter two are consistent with previously reported values in cell medium for PCL10_PEG10 and PCL20_PEG10.[42] Mean stiffness values are 1.7 and 1.5 N/m for PCL10_PEG2 and PCL20_PEG5, respectively, while again PCL20_PEG10 and PCL10_PEG10 display much softer properties by an order of magnitude of 0.3 and 0.2 N/m, respectively. The mean values of the adhesion force are 0.8, 0.4, 0.5 and 0.1 nN for PCL10_PEG2, PCL20_PEG5, PCL20_PEG10 and PCL10_PEG10, respectively, overall decreasing slightly with increasing PEG content.

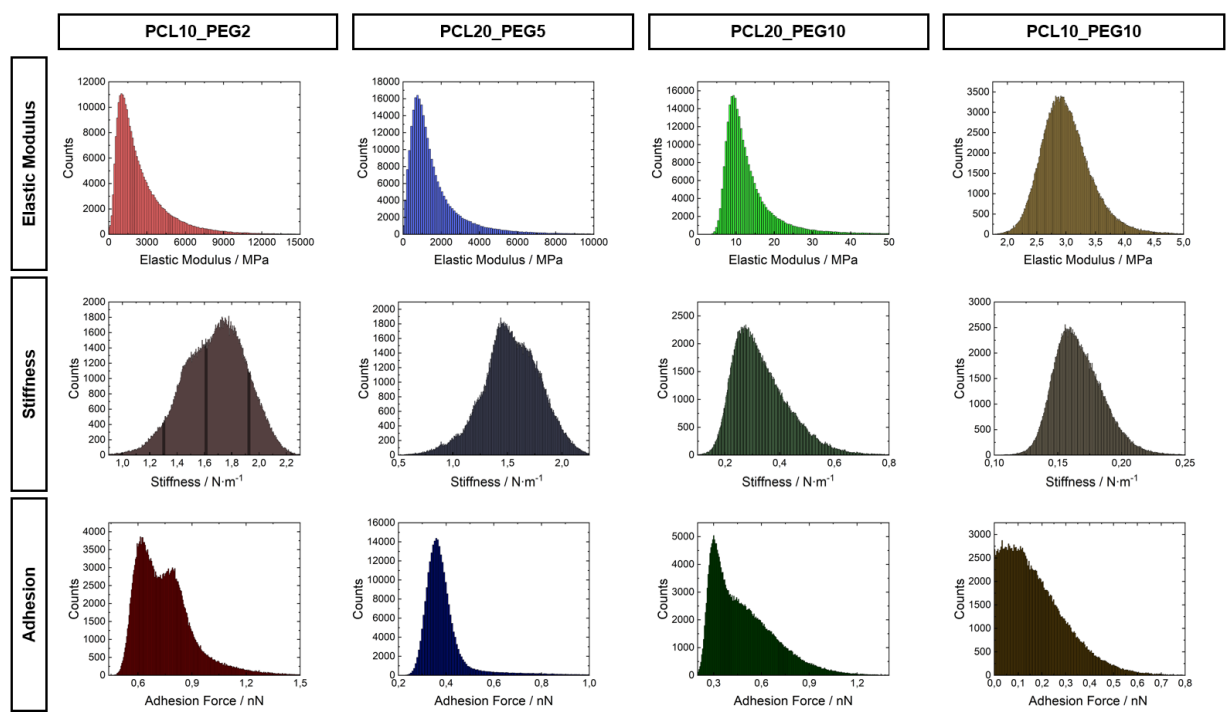

**Figure 8:** Atomic force microscopy – Quantitative imaging. Quantitative results for elastic modulus (row 1), stiffness (row 2) and adhesion force (row 3) for PCL_PEG films with varying star size in water.

Figure 10 illustrates image data derived from these measurement modes in the non-selective solvent toluene. It has previously been demonstrated that PCL_PEG networks swollen in the non-selective solvent toluene display no structure on a nanoscale and are consequently not of primary interest in this study (Figure 5).[38] Therefore, toluene films are rather analyzed on the micrometer scale, and the image size is increased to 10 x 10 µm$^2$. The first row in Figure 10 shows the height topography at the contact point during the indentation approach. None of the samples display preferential orientation of microstructures. PCL10_PEG2 shows connected patchy areas, while PCL20_PEG5 and PCL20_PEG10 exhibit a connected amorphous network, and PCL10_PEG10 shows islands connected by border-valleys, although the overall structure appears amorphous. It has been demonstrated in previous works that these islands are a result of the spin coating procedure, and they persist for both water and toluene swollen systems.[42] The sample roughness shows no correlation with composition and ranges from 23 to 45 nm, consistent with measurements for thin films in previous studies.[42]

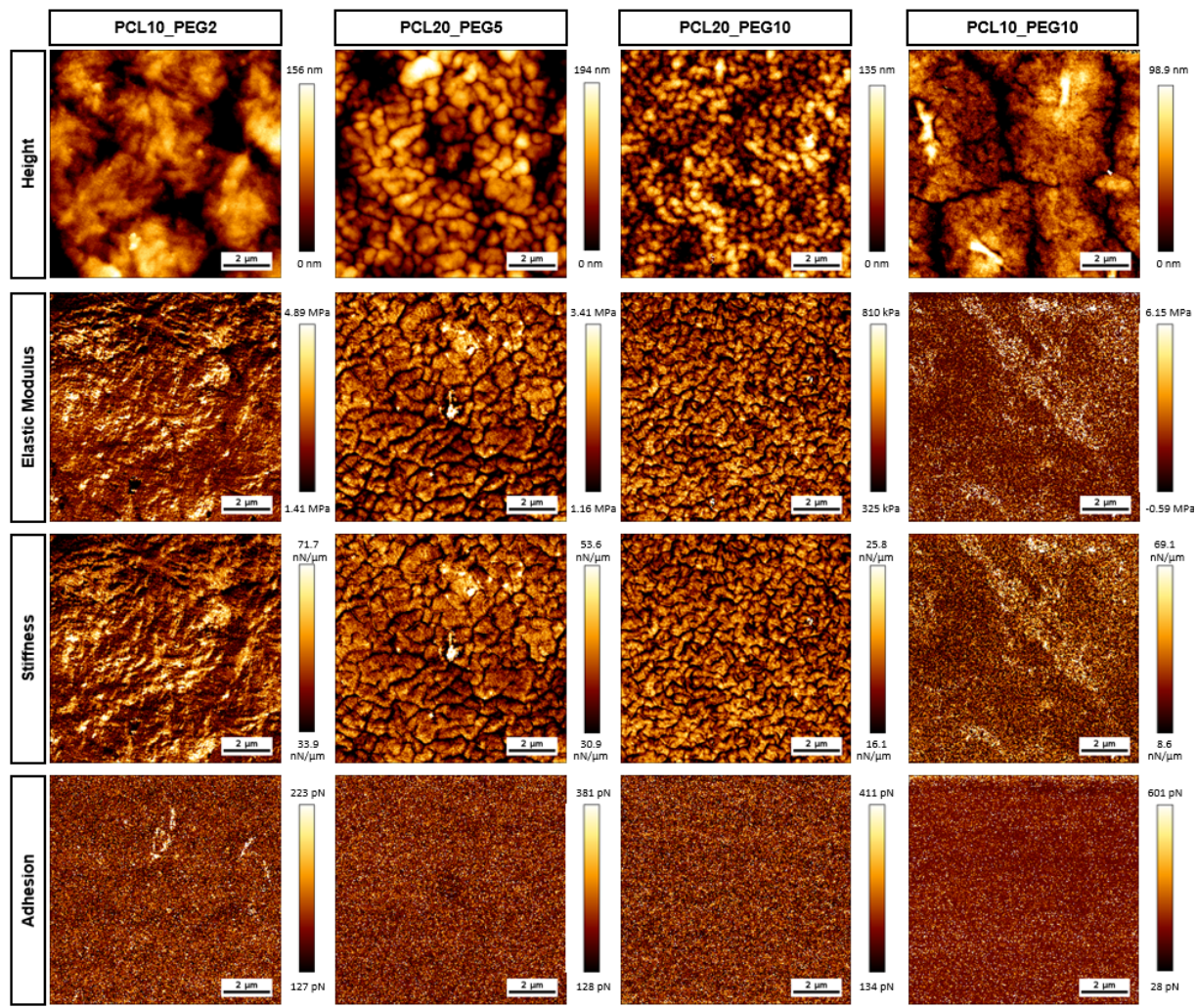

**Figure 9:** Atomic force microscopy - quantitative imaging of PCL_PEG thin films with varying star size in toluene. First row: Height image based on the contact point. Second row: Elastic moduli based on Hertz fit. Third row: Sample stiffness based on tangential fit. Fourth row: Adhesion force. Image size = 10 x 10 µm$^2$. Scale bar = 2 µm.

The second and third row of Figure 10 display the elastic modulus and stiffness, respectively. Since these images are not related to an absolute height value, but rather a sample property, structural features are easier to highlight and verify. The PCL10_PEG2 film appears to be structurally uniform on a microscopic scale, while the PCL20_PEG5 and PCL20_PEG10 films retain their interconnected network already displayed in the height images. Interestingly, the PCL10_PEG10 film (highest PEG content) is most uniform across the entire sample surface. Adhesion imaging (Figure 10, row 4) displays the lowest contrast and highest uniformity of all types of AFM maps, and it is similar irrespective of the PCL/PEG composition. This indicates that the network composition has no significant influence on the adhesion properties in non-selective solvents.

All measured properties are evaluated quantitatively. Results obtained for elastic modulus, stiffness and adhesion are displayed in Figure 11. Numeric values, including mean and standard deviation, are displayed in the supporting information Figure S2. Elastic moduli in toluene are 2.4, 1.8, 0.5 and 1.3 MPa for PCL10_PEG2, PCL20_PEG5, PCL20_PEG10 and PCL10_PEG10, respectively. The results indicate that the elastic modulus decreases with increasing network precursor size. The same trend is observed for the corresponding stiffness values, ranging from 0.04 to 0.02 N/m for all samples. As already highlighted in the image analysis, adhesion forces are similar for all sample types and are very low, ranging from 0.15 to 0.21 nN. In general, the adhesion in toluene is lower than in water.

To summarize, the chemical composition has much less effect on the physical features of the surfaces in toluene than in water.

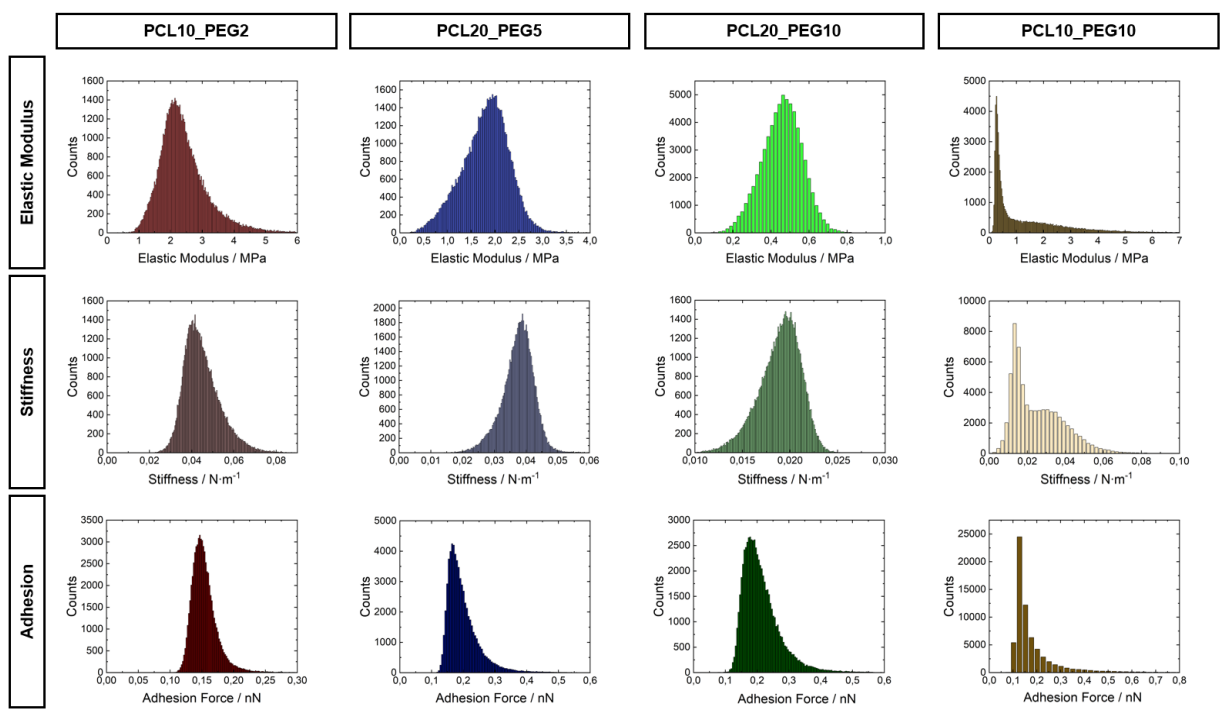

**Figure 10:** Atomic force microscopy – Quantitative imaging. Quantitative results for elastic modulus (row 1), stiffness (row 2) and adhesion force (row 3) for PCL_PEG films with varying star size in water.

*Viscoelastic properties of APCN in water and toluene*

All sample types are analyzed in terms of their rheological properties. The samples are swollen in either water or toluene and subsequently undergo dynamic indentation experiments with a nanoindenter. Nanoindentation is performed at various indentation forces, ranging from 5 – 20 nN. Once the indenter experiences its setforce, it oscillates at a given frequency before retracting. Oscillations are performed at a frequency range of 10 – 250 Hz in logarithmic steps. Resulting oscillation curves are fitted with the Hertz-Model and yield the storage modulus E', the loss modulus E'' and the loss tangent (E''/E').

In the first set of experiments, all samples were swollen in water (Figure 12). For all sample types, both the storage modulus E' and loss modulus E'' increase with increasing setforce (5 nN, 10nN and 20 nN). Since an increase in setforce is directly related to a deeper maximum indentation, we can formulate that the moduli increase with larger indentation depths. In addition, both storage and loss moduli increase monotonically with applied oscillation frequency. Both storage and loss modulus decrease with increasing PEG content, which has already been indicated by the static indentation measurements shown above. For all sample types, the storage moduli exhibit a weak positive slope on the log–frequency axis. At the lowest setforce (5 nN), they rose from roughly 0.4 - 1.0 MPa at 10 Hz to 0.45 - 1.5 MPa near 250 Hz. Increasing the load to 10 nN and 20 nN shifted the entire modulus-frequency curves upward by approximately 50 - 70 % and 40 - 100 %, respectively, while preserving the curve shape. At a setforce of 10 nN, storage moduli range from 0.6 - 1.5 MPa at the lowest frequency and 0.7 - 1.7 MPa at the highest frequency. In comparison, a setforce of 20 nN results in storage moduli of 0.9 - 2.2 MPa (at 10 Hz) and 1.1 - 3.0 MPa (at 250 Hz). Similar trends were observed for the loss modulus but with lower absolute magnitudes. At 5 nN, the loss moduli increased from about 0.03 - 0.25 MPa (10 Hz) to 0.45 - 0.30 MPa (250 Hz), depending on sample type. At 10 nN and 20 nN, the curves are again vertically shifted upward while preserving a moderate slope, with loss moduli ranging from 0.05 - 0.25 MPa (10 Hz) to 0.08 - 0.35 MPa (250 Hz) and 0.08 - 0.4 MPa (10 Hz) to 0.13 - 0.8 (250 Hz), respectively. When comparing the different star types, both storage and loss modulus decrease with increasing PEG content. This is a result of networks with higher PEG content showing a larger ability to swell in the selective solvent water. For all sample types, the storage modulus E' is roughly one order of magnitude larger than the loss modulus E''. This is highlighted in Figure 12 by displaying the loss tangent E''/E' as a function of frequency. All loss tangents increase monotonically with frequency, thereby highlighting that energy dissipation is becoming a more dominant factor at high frequencies. In the frequency range of 10 – 250 Hz, the loss tangent increased from 0.18 to 0.26 for PCL10_PEG2, 0.15 to 0.18 for PCL20_PEG5, 0.10 to 0.14 for PCL20_PEG10 and 0.08 to 0.12 for PCL10_PEG10. Since the loss tangent represents the ratio of viscous and elastic contributions, the low loss tangents in all observed samples clearly highlight that PCL_PEG gel films in this study are primarily dominated by elastic behavior and are therefore rather solid- than liquid-like. Whereas the slope of the loss tangent curves increases linearly for PCL10_PEG2 and PCL20_PEG5, the slope of PCL20_PEG10 and PCL10_PEG10 increases parabolically, with higher frequencies noticeably increasing the loss tangent. Interestingly, the loss tangent decreases with increasing swellability of the films. Further, the loss tangent shows no significant difference with the applied setforce. The viscoelastic behavior appears mostly independent of the tested indentation depth range, which means that stress-strain relationships and energy dissipation scale proportionally. This would suggest a rather homogeneous polymer network, at least on the investigated length scale.

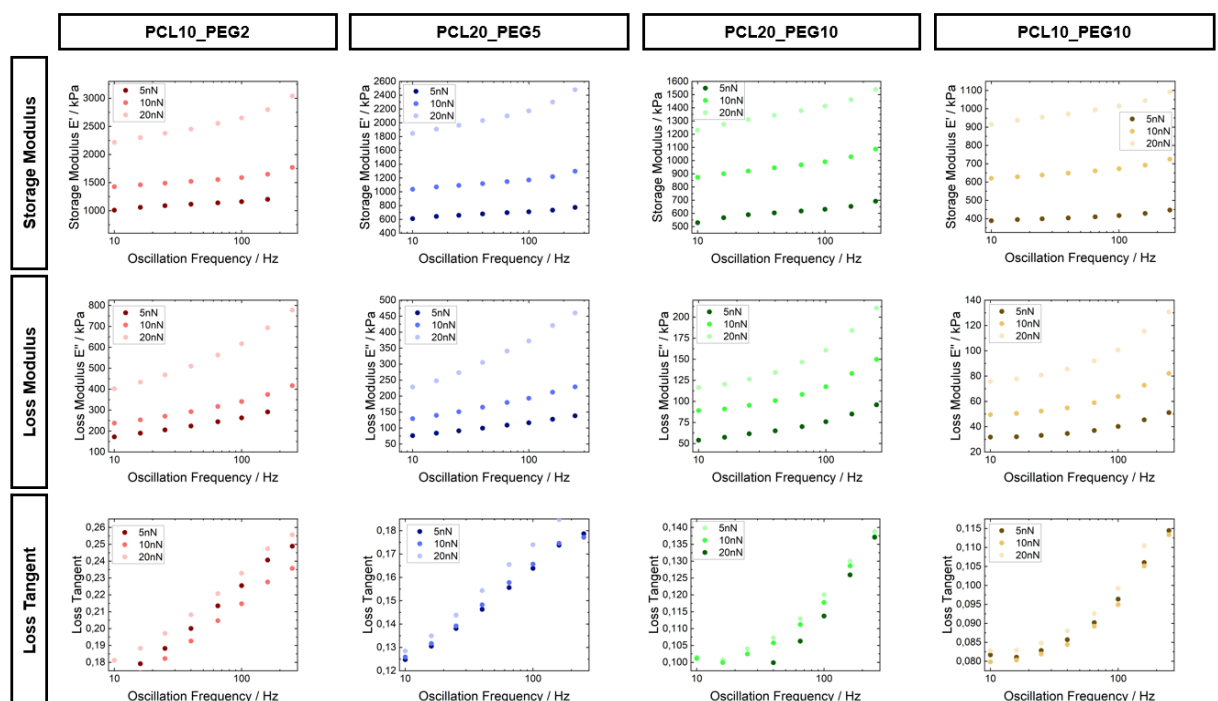

**Figure 11:** Nano-rheological behavior of the surface of PCL_PEG films in water obtained from dynamic indentation measurements with AFM. Storage modulus E' (top row), E'' (middle row), loss tangent (E''/E') as a function of oscillation frequency (10 – 250 Hz). Error bars are omitted for visual clarity. Error bars consistently range from 10 – 25% of the displayed average values for all samples.

In the second set of experiments, all sample types were swollen in the non-selective solvent toluene and analyzed in terms of their rheological properties for set forces of 1, 2 and 5 nN (Figure 13). A lower setforce was chosen than in water, since the films are softer in toluene than in water, as shown above for the static indentation measurements. All moduli are lower than in water due to the stronger swelling ability of the PCL_PEG films in toluene than in water. The storage moduli E' increase only slightly with oscillation frequency. At the lowest setforce (1 nN), they range from 0.05 - 0.25 MPa (10 Hz) to 0.05 - 0.30 MPa (250 Hz) depending on sample type. They increase from roughly 0.4 - 1.0 MPa at 10 Hz to 0.45 - 1.5 MPa near 250 Hz. Increasing the load to 2 nN and 5 nN shifted the entire frequency curves upward by approximately 30 - 60 % and 45 - 90 %, respectively, while preserving the curve shape. At a setforce of 2 nN, storage moduli range from 0.06 - 0.40 MPa at the lowest frequency and 0.06 - 0.5 MPa at the highest frequency. In contrast, a setforce of 20 nN results in storage moduli of 0.1 - 0.7 MPa and 0.11 - 0.9 MPa for frequencies from 10 to 250 Hz. The loss modulus E″ overall displays lower magnitudes and shows slightly different trends. While the overall curves move parabolically upwards, they seem to be almost constant for low frequencies between 10 - 50 Hz and then slowly increase up to 250 Hz. At 1 nN, the loss moduli increased from about 0.01 - 0.015 MPa (10 Hz) to 0.01 - 0.02 MPa (250 Hz) depending on sample type. At 2 nN and 5 nN, the curves are again vertically shifted upward while preserving the curve shape, with loss moduli ranging from 0.01 - 0.02 MPa (10 Hz) to 0.01 - 0.03 MPa (250 Hz) and 0.01 - 0.04 MPa (10 Hz) to 0.01 - 0.06 (250 Hz), respectively. In contrast to water swollen films, films which are swollen in toluene do not show a monotonous decrease in storage and loss modulus and loss tangent with increasing PEG content. PCL20_PEG10 films show the lowest storage and loss modulus, which indicates an effect of the star size on the rheological properties. Obviously, networks with larger stars show a larger ability to swell in the non-selective solvent toluene. All sample types display a storage modulus E' roughly one order of magnitude larger than the loss modulus E" as in water. In contrast to films swollen in water, the loss tangents E"/E' are mostly constant at low frequencies (roughly 10 – 50 Hz) and increase afterwards up to 250 Hz. This indicates that smaller perturbations to the network cause a constant viscoelastic response, while higher frequencies favor energy dissipation. In the frequency range 10 - 250 Hz, loss tangents range from 0.05 - 0.08 for PCL10_PEG2, 0.05 - 0.10 for PCL20_PEG5, 0.04 - 0.11 for PCL20_PEG10, and 0.06 - 0.08 for PCL10_PEG10. While the loss tangent for

PCL_10_PEG2 is independent of the setforce and therefore indentation depth, there seems to be a scaling of loss tangent with higher indentation depths, which is more pronounced for lower frequencies and diminishes for higher frequencies. While this may suggest a more complex viscoelastic response for fully swollen networks with larger star sizes, the absolute difference is marginal (maximum difference between loss tangent measurements at 1 nN and 5 nN is 40%, while the force increase is 400%), further indicating that the networks in this study behave rather homogeneously.

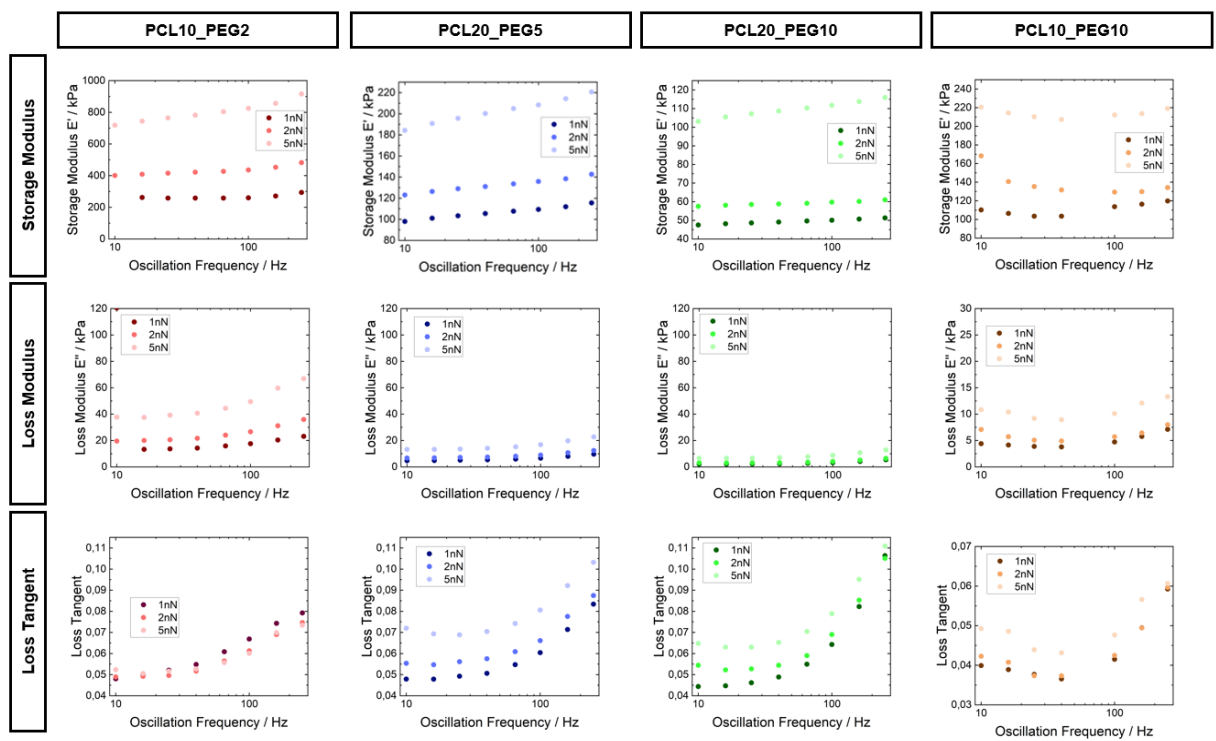

**Figure 12:** Nano-rheological behavior at the surface of PCL_PEG films in toluene obtained from dynamic indentation measurements with AFM. Storage modulus E' (top row), E'' (middle row), loss tangent (E''/E') as a function of oscillation frequency (10 – 250 Hz). Error bars are omitted for visual clarity. Error bars consistently range from 10 – 25 % of displayed average values for all samples.

*Temperature-dependent elastic moduli and dynamic crystallization effects*

To study the effect of temperature on the structure and mechanical properties of PCL_PEG co-networks in the selective solvent water, samples were subjected to a temperature ramp. Monitoring via AFM quantitative imaging techniques yielded structural information as well as elastic moduli distributions down to the nanoscale. Thin PCL_PEG films in water were first heated up from 25 °C to 65 °C in 5°C-steps, exceeding the melting point of the networks in bulk. At each temperature increment, samples were left to equilibrate for at least 15 min and then measured. After reaching the final temperature of 65 °C, the process was run in reverse. Samples were cooled in 5 °C increments down to 25 °C, while equilibrating in the same manner and monitoring via AFM quantitative imaging. Figure 14 displays the elastic modulus in dependence of the applied temperature during the ramp-up and down cycle. Samples with high PCL content, PCL10_PEG2 and PCL 20_PEG5, start with high elastic moduli of 600 MPa and 650 MPa at 25 °C, due to the incompatibility of PCL and the selective solvent water. During the temperature ramp-up cycle, the elastic modulus decreases until it drops sharply at the estimated melting point of the polymer network and then stays at constantly low values of about 1 MPa for both sample types. The shape of the curve is sigmoidal, and the melting point of the thin film network is therefore estimated from the point of deflection. For PCL10_PEG2, this transition occurs at 45 °C, and for PCL20_PEG5, it occurs at 47°C. During the subsequent cooldown cycle, both sample types maintain their low elastic moduli beyond the estimated melting point or presumed re-crystallization point of the network. The moduli remain at roughly 1 MPa until 35 °C. However, they increase abruptly when cooling below 30°C. At 25 °C, the resulting elastic moduli are 200 MPa for PCL10_PEG2 and 100 MPa for PCL20_5. This is significantly lower than the elastic moduli at the start of the experiment. The overall mechanical behavior shows a strong hysteresis. Co-networks with higher PEG content, PCL20_PEG10 and PCL10_PEG10, also display a strong hysteresis, initially displaying a similar ramp-up behavior. They start with significantly lower elastic moduli of 50 MPa for PCL20_PEG10 and 110 MPa for PCL10_PEG10 at 25 °C due to stronger swelling in water because of the higher PEG content. Elastic moduli decrease with increasing temperature, and the transition temperature occurs at 52°C for PCL20_PEG10 and 45 °C for PCL10_PEG10. Beyond the melting point, the elastic moduli are roughly 1 MPa. During the entire cooling process, the samples maintain their low elastic moduli around 1 MPa. If samples are left for multiple weeks, they will keep crystallizing and regain their initially high elastic moduli, although this process is slower the higher the PEG content. Additional structural information during the ramp cycles on exemplary samples can be found in the supporting information and convincingly demonstrates how changes in mechanical properties coincide with structural changes (Figure S1 and S2).

AFM micrographs of the surfaces during the heating/cooling cycle are shown in Figure S1 and S2.

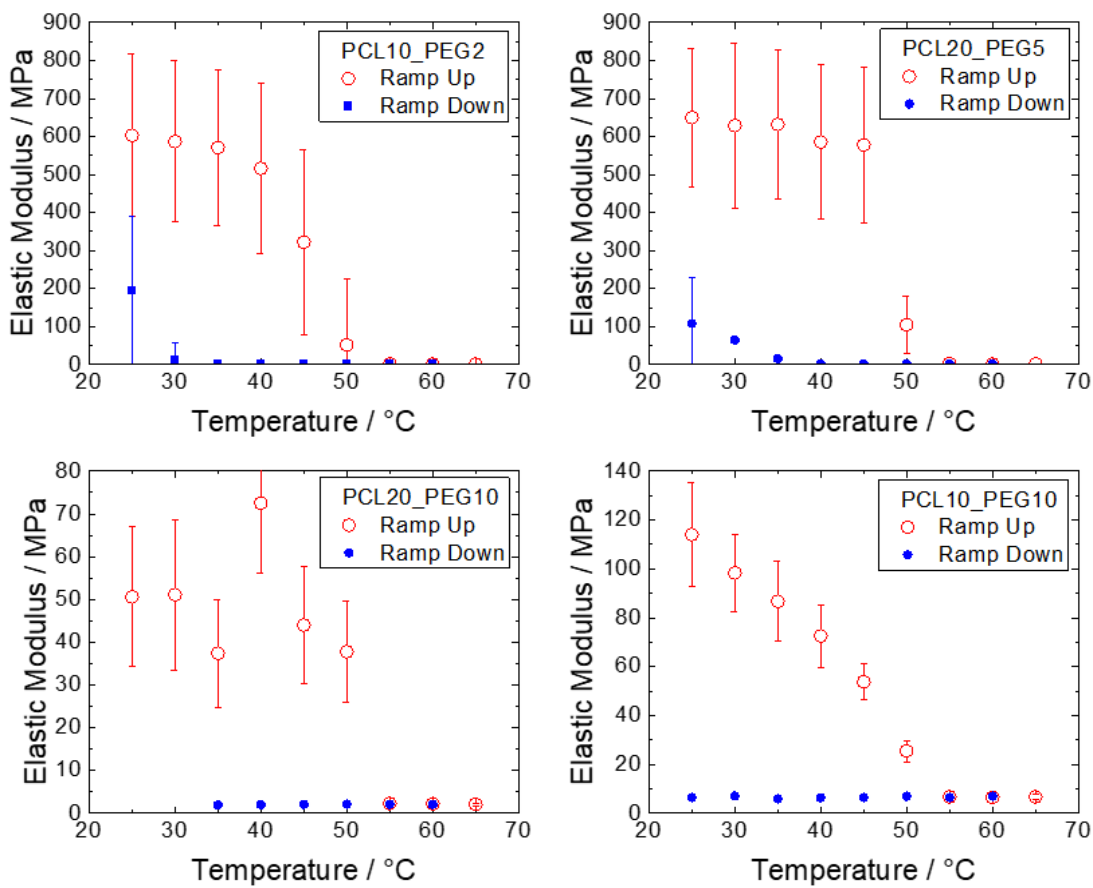

**Figure 13:** Elastic Moduli of PCL_PEG films in water during a temperature ramp cycle.

**Discussion**

The present study addresses films of amphiphilic polymer co-networks (APCNs) composed of tetra-PEG and tetra-PCL co-networks. Specific focus is on the effect of the size of the network precursors, their combination and the surrounding solvent on the network architecture of the films and their mechanical, rheological and adhesive properties. Emphasis is placed on employing the cutting-edge AFM techniques to evaluate structure-property relationships down to the nanoscale.

AFM phase imaging techniques and the use of ultra-sharp cantilevers revealed that PEG_PCL co-networks exhibit nanoscopic phase separation in selective aqueous conditions (Figure 6 and 7), as expected due to the incompatibility between hydrophobic PCL and water, leading to clustering of PCL.[38]

It is known that spherical, cylindrical, lamellar and related structures exist in amphiphilic co-networks and block-copolymer systems depending on the polymer volume fractions.[44],[45,46]

For the studied system networks with higher PCL content (PCL10_PEG2, PCL20_PEG5 and PCL20_PEG10) display semi-crystalline, cylindrical nanostructure arrangements with preferential orientation. In contrast, samples with higher PEG content (PCL10_PEG10) appeared amorphous with spherical nanophase domains, as has been reported previously for bulk samples (Figure 6 and 8).[38] This transition from spherical to cylindrical structures suggests that increasing hydrophobic star size beyond some critical ratio promotes semicrystalline arrangements, likely driven by PCL crystallization in the selective solvent.

Systems composed of pure PCL crystallize in the form of spherulites in confined systems,[47,48] favoring cylindrical alignment in tetra-PCL rich co-networks. Mixed forms of cylindrical and spherical nanophase separation have been observed in samples with higher PCL content (Figure 6 and 7). On one hand, this could be a result of cylindrical structures facing the sample surface, thereby appearing spherical. On the other hand, the mixed structures in these co-networks could heavily depend on the dynamics of the crystallization process. This is highlighted by a strong hysteresis in temperature-dependent indentation measurements (Figure 14), as well as dramatic increases in elastic moduli up to hundreds of MPa (Figure 9) for PCL-rich samples when stored in ambient conditions.

Quantitative AFM measurements demonstrate great differences in elastic modulus, stiffness and to some degree adhesion in all PCL_PEG films in selective (water) and non-selective (toluene) conditions (Figure 9 and 11). In water, thin films exhibited elastic moduli up to several hundred MPa or even GPa for PCL-rich networks. In contrast, PEG-rich networks exhibit significantly lower moduli of only a few MPa. The elastic modulus decreases monotonously with increasing PEG content. This highlights the inability of PCL to swell in water, which restricts the expansion of the network and enhances its stiffness. The simultaneous determination of structure and mechanics down to a few nanometers allows for the direct correlation of structural features and mechanical properties. Crystallization in PCL_PEG networks significantly increases the elastic moduli by orders of magnitude, most notably for PCL-rich samples. Apart from the lower swelling ability of PCL-rich networks, the structural arrangement of PCL in cylindrical structures may also contribute to higher elastic moduli and stiffness.

In contrast to the high moduli observed for aqueous systems, all samples swollen in toluene showed much lower moduli of a few hundred kPa to low MPa range, indicating that the non-selective solvent toluene effectively swells both network constituents, consistent with results for bulk swelling experiments.[38] Elastic moduli and stiffness in these networks clearly show a trend to lower values as the overall size of star precursors increases (Figure 11). Unlike observations in water (Figure 9), PCL10_PEG10 is no longer the softest sample, but rather PCL20_PEG10, which has a greater ability to swell simply due to the length of its tetra-arms. It is clear from this result that mechanical differences between tetra-PCL and tetra-PEG are eliminated in non-selective solvents. This leaves only microstructures to be analyzed for samples in toluene, which are likely a result of the initial spin coating process, where early solvent removal forces different levels of the network to interconnect, creating tension and structural deformation. This was demonstrated by an amorphous surface structure with lots of gaps and valleys (Figure 10).

The adhesion forces on a nanoscale appear to be low overall and only show a slight correlation with co-network type in the case of water-swollen systems. The adhesion forces decrease slightly with an increase in PEG content, potentially indicating that non-swollen PCL content makes the networks stickier. This is evident for films with high PCL content (Figure 8, first and second column), where the objects of a size of a few 10 nm, detected in the height image, show a high adhesion force. These objects are interpreted as PCL excess phases. They are not detectable in the maps of the elastic modulus and the stiffness. Obviously, they act as an extended tip during indentation and are pushed into the softer polymer matrix.

In toluene, PCL and PEG become indifferent, overall displaying a homogeneous sample surface in terms of adhesion, while the presence of surface structures and large surface roughness might suggest otherwise.

Dynamic indentation experiments show that all network types are predominantly exhibiting elastic behavior, rather than viscous behavior, with storage moduli (E′) exceeding loss moduli (E″) by approximately one order of magnitude for the tested frequency range of 10 - 250 Hz (Figure 12 and 13). For co-networks in water, storage and loss moduli decrease with increasing PEG content, whereas for co-networks in toluene, storage and loss moduli decrease with an increase in the size of star precursors within a network. Loss tangents increase with frequency, showing that energy dissipation is a more dominating factor compared to elastic storage on short time scales. In aqueous conditions, PEG-rich networks exhibited more pronounced frequency-dependent viscoelasticity, with a tendency towards greater energy dissipation at higher deformation rates. This is likely due to increased water uptake and the resulting enhanced chain mobility. Against our expectation, the loss tangent decreases with increasing PEG content. This might be explained by a stronger friction of PCL crystals in the PCL-rich films. All sample types in both solvent conditions show loss tangents mostly independent of indentation depth. This suggests that the co-networks are structurally homogeneous at the probed scales. This statement is confirmed by the presence of structural information from various imaging techniques. While phase separation is detectable at the interface, these structural patterns continue in the bulk phase underneath the surface. This observation is particularly emphasized by the similarities of structural patterns in stiffness imaging at larger indentation depths (Figure 8) and phase imaging representing solely superficial structures (Figure 6 and 7).

Temperature-dependent AFM measurements demonstrated a strong hysteresis for all samples in aqueous conditions (Figure 14). The initially high moduli are likely a result of co-networks stored in ambient conditions, allowing crystallization processes to proceed. Elastic moduli decrease sigmoidally with rising temperatures, with transition temperatures near 45 °C, which corresponds to tetra-PCL melting temperature. A strong shift of the melting/crystallization temperature from roughly 45 °C during heating, to less than 30 °C during cooling in the PCL-rich networks was observed. This indicates that PCL crystallization is strongly dependent on network kinetics, where soluble PEG components prevent PCL reorganization. For networks with higher PEG content (PCL20_PEG10, PCL10_PEG10) crystallization is completely prevented on the observed time scale of the experiment. Given the high moduli starting conditions, however, it can be concluded that sample storage in ambient conditions allows crystallization processes to proceed, even for PEG-rich networks.

The observed structure-mechanics relationships highlight the control of APCNs through co-network composition and solvent environment. Irrespective of their composition, the studied co-networks have in common that they exhibit low adhesion, primarily elastic properties with frequency-dependent viscoelasticity and large hysteresis during heating/cooling cycles.

Large PCL content in tetra-PCL_tetra-PEG co-networks in water-based systems leads to semi-crystalline microstructures and cylindrical nanostructures, which present stiff networks. Large PEG content in tetra-PCL_tetra-PEG co-networks in aqueous environments leads to amorphous microstructures, spherical nanophase separation and absence of crystallization when carefully choosing the environment. The films are softer with a more pronounced frequency-dependent viscoelasticity at high deformation rates. Some structural and property differences between PCL_films of different composition are eliminated when using non-selective solvents. However, control over PCL and PEG star size, PCL/PEG ratio and synthesis conditions still governs the resulting microstructural patterns and mechanical properties of the co-networks, with large star precursors being the main contributor towards softer, primarily elastic gels with uniform adhesion and deformation-rate-dependent viscoelastic properties.

**Conclusion**

This work demonstrates the development of a predictable nanoscopic structure-property relationship for films made from amphiphilic polymer co-networks based on tetra-PCL and tetra-PEG with varying star sizes.

It highlights the potential of advanced atomic force microscopy techniques to create outstanding structural contrasts while simultaneously monitoring a large spectrum of sample properties like nanomechanics, nanorheology and adhesion. The results show a strong correlation between the properties of the gel films and their structures on the nanoscale. The network architecture is adjustable based on the selection of suitably sized precursor tetra-armed stars. The resulting elastic moduli range from a few kPA to hundreds of MPa, depending on the gel composition, environmental conditions, such as solvent type and temperature. The networks are predominantly elastic and display frequency-dependent viscoelasticity with higher amounts of energy dissipation at large deformation rates (i.e. short time scales). Selective hydrophilic solvents like water cause a transition from nanoscopic spherical to cylindrical phase separation and strong increase in the elastic modulus and stiffness by increasing the total amount of hydrophobic PCL in the network. Temperature-dependent properties show a large hysteresis and highlight the kinetics of the crystallization process in the constant battle of PCL crystallization and PEG solubilization. It also demonstrates the importance of sample history and measurement conditions.

These valuable insights create actionable templates for designing surface adaptive materials with desired structure-property outcomes. Modern AFM measurements provide a simple, fast and powerful technique to guide this process.

**Funding:** This study was conducted as part of the research collaboration "Adaptive Polymer Gels with Model-Network Structure" (FOR2811), funded by the German Research Foundation (DFG), grant 397384169, along with grants 423768931 (TP5) and 423514254 (TP1).

**Conflicts of Interest:** The authors declare no conflict of interest.

**Supporting information**

**Table S1**: APCN thin film properties in water

| Sample property / unit | PCL10_PEG2 | PCL20_PEG5 | PCL20_10 | PCL10_PEG10 |
|---|---|---|---|---|
| Elastic modulus / MPa | 2500 ± 2000 | 1600 ± 1400 | 13.1 ± 6.6 | 3.0 ± 0.4 |
| Stiffness / N m$^{-1}$ | 1.70 ± 0.20 | 1.50 ± 0.26 | 0.33 ± 0.10 | 0.17 ± 0.02 |
| Adhesion / nN | 0.75 ± 0.15 | 0.39 ± 0.09 | 0.49 ± 0.20 | 0.09 ± 0.17 |

**Table S2**: APCN thin film properties in toluene

| Sample property / unit | PCL10_PEG2 | PCL20_PEG5 | PCL20_10 | PCL10_PEG10 |
|---|---|---|---|---|
| Elastic modulus / MPa | 2.4 ± 0.8 | 1.8 ± 0.5 | 0.5 ± 0.1 | 1.3 ± 1.3 |
| Stiffness / N m$^{-1}$ | 0.04 ± 0.01 | 0.04 ± 0.01 | 0.02 ± 0.01 | 0.03 ± 0.01 |
| Adhesion / nN | 0.15 ± 0.02 | 0.20 ± 0.05 | 0.21 ± 0.06 | 0.19 ± 0.13 |

Next to the sample properties, structural changes during the ramp cycles were monitored. Figure S1 and S2 demonstrate the structural changes of PCL10_PEG10 (highest PEG ratio) and PCL10_PEG2 (lowest PEG ratio). Even PCL10_PEG10, despite its overall amorphous appearance, shows signs of defined structuring at the start of the experiment at 25 °C (Figure S1). After heating beyond the melting point of the network, moduli drop dramatically, while previously distinct structuring disappears. Moduli do not increase after the cool down and the sample remains amorphous.

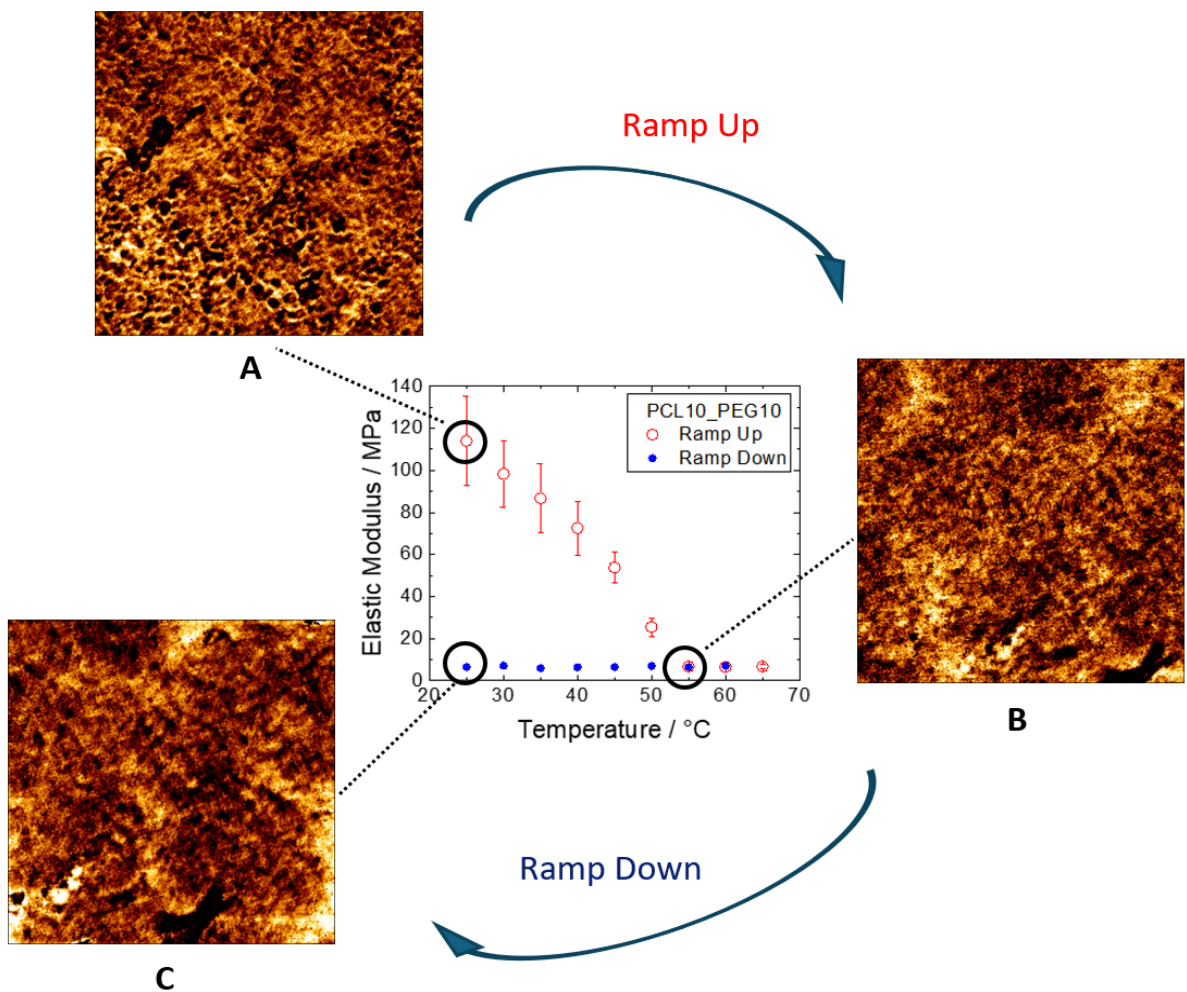

**Figure S1:** Surface topography and elastic modulus during temperature cycle for PCL10_PEG10. (A) 25 °C, (B) 55 °C, (C) 25 °C. Image size 5 x 5 µm$^2$.

Contrarily, PCL10_PEG2, shows strong signs of semi-crystallinity in the form of spherulites at the start of the experiment at 25 °C, with clear distinction of locally ordered nano-cylinders (Figure S2). After heating the sample to 65 °C, spherulitical structures are still visible. However, the local ordering is no longer recognizable. This is accompanied by a steep drop in elastic moduli. After the sample is cooled down to 25 °C, elastic moduli rise, and distinct semi-crystalline structures reappear.

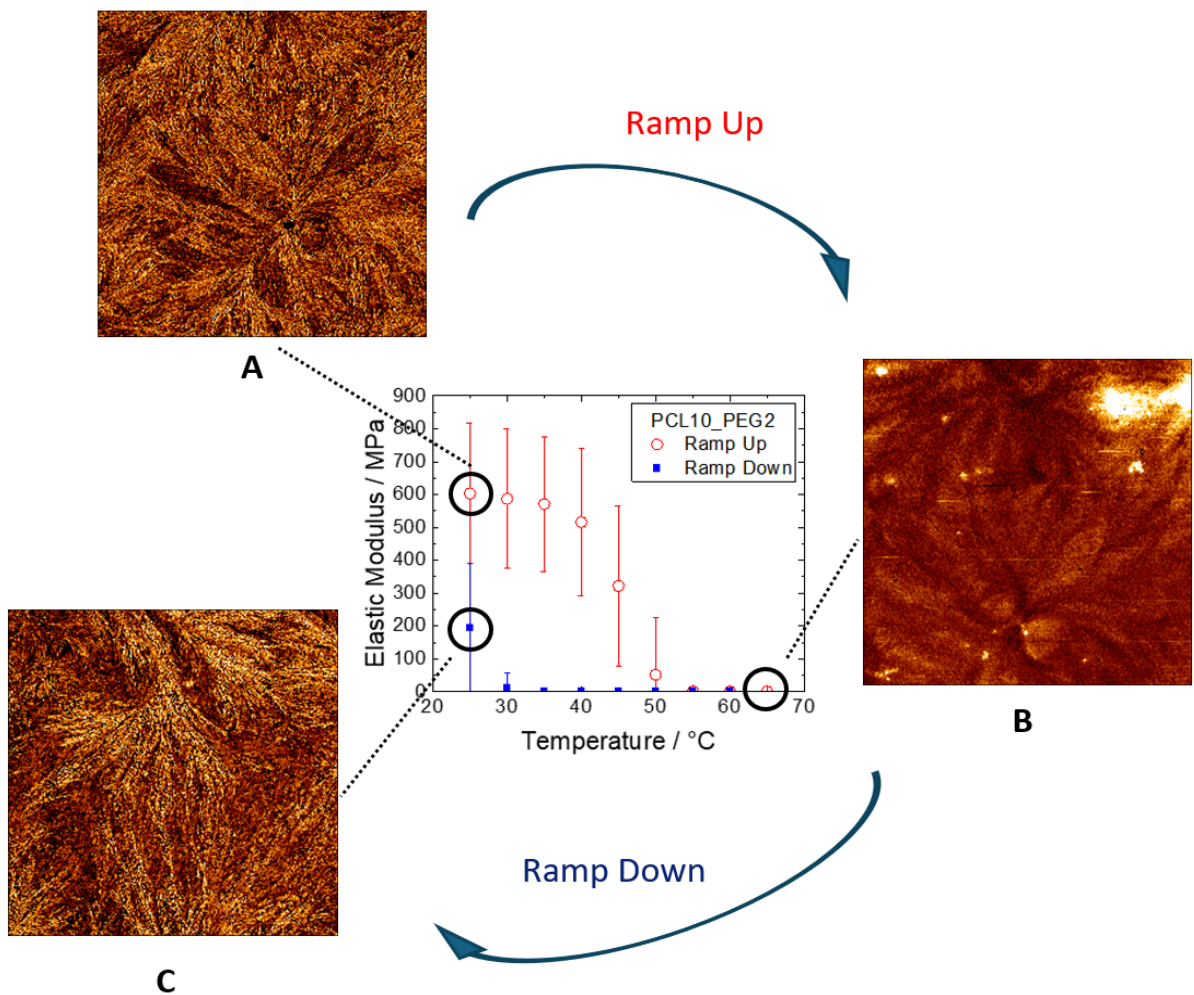

**Figure S2:** Surface topography and elastic modulus during temperature cycle for PCL10_PEG2. (A) 25 °C, (B) 65 °C, (C) 25 °C. Image size 5 x 5 µm$^2$.